\documentclass[sigconf]{acmart}
\AtBeginDocument{%
  \providecommand\BibTeX{{%
    \normalfont B\kern-0.5em{\scshape i\kern-0.25em b}\kern-0.8em\TeX}}}

\copyrightyear{2024}
\acmYear{2024}
\setcopyright{acmlicensed}\acmConference[WWW '24 Companion]{Companion
Proceedings of the ACM Web Conference 2024}{May 13--17, 2024}{Singapore,
Singapore}
\acmBooktitle{Companion Proceedings of the ACM Web Conference 2024 (WWW
'24 Companion), May 13--17, 2024, Singapore, Singapore}
\acmDOI{10.1145/3589335.3651558}
\acmISBN{979-8-4007-0172-6/24/05}

\usepackage{colortbl}
\usepackage{multirow}

\begin{document}

\title{Discrete Semantic Tokenization for Deep CTR Prediction}

\author{Qijiong Liu}
\authornote{The work was done when the author was a visiting student at NUS.}
\affiliation{%
  \institution{PolyU, Hong Kong}
  \country{}
}
\email{liu@qijiong.work}

\author{Hengchang Hu}
\affiliation{%
  \institution{National University of Singapore}
  \country{}
}
\email{hengchang.hu@u.nus.edu}

\author{Jiahao Wu}
\affiliation{%
  \institution{PolyU, Hong Kong}
  \country{}
}
\email{jiahao.wu@connect.polyu.hk}

\author{Jieming Zhu}
\affiliation{%
  \institution{Huawei Noah's Ark Lab, China}
  \country{}
}
\email{jiemingzhu@ieee.org}

\author{Min-Yen Kan}
\authornote{Min-Yen Kan and Xiao-Ming Wu are the corresponding authors.}
\affiliation{%
  \institution{National University of Singapore}
  \country{}
}
\email{kanmy@comp.nus.edu.sg}

\author{Xiao-Ming Wu}
\authornotemark[2]
\affiliation{%
  \institution{PolyU, Hong Kong}
  \country{}
}
\email{xiao-ming.wu@polyu.edu.hk}

\renewcommand{\shortauthors}{Qijiong Liu, Hengchang Hu, Jiahao Wu, Jieming Zhu, Min-Yen Kan, \& Xiao-Ming Wu}
\newcommand{\model}{UIST}
\newcommand{\smodel}{IST}

\newcommand{\kmy}[1]{\textcolor{magenta}{$_{min}$[#1]}}

\begin{abstract}
Incorporating item content information into click-through rate (CTR) prediction models remains a challenge, especially with the time and space constraints of industrial scenarios. The content-encoding paradigm, which integrates user and item encoders directly into CTR models, prioritizes space over time. In contrast, the embedding-based paradigm transforms item and user semantics into latent embeddings, subsequently caching them to optimize processing time at the expense of space. In this paper, we introduce a new semantic-token paradigm and propose a discrete semantic tokenization approach, namely \model{}, for user and item representation.  \model{} facilitates swift training and inference while maintaining a conservative memory footprint. Specifically, \model{} quantizes dense embedding vectors into discrete tokens with shorter lengths and employs a hierarchical mixture inference module to weigh the contribution of each user--item token pair. Our experimental results on news recommendation showcase the effectiveness and efficiency (about 200-fold space compression) of \model{} for CTR prediction.
\end{abstract}




\maketitle

\section{Introduction}

Click-through rate (CTR) prediction models~\cite{FuxiCTR} 
aim to predict the probability of users interacting with items. The real-time demands of online services~\cite{liu2022prec,liu2023only} pose a significant challenge in seamlessly merging such deep CTR models with valuable semantic knowledge, encompassing item content and user history.

Traditional deep CTR models commonly depend on ID-based approaches, incorporating features like item and user IDs, along with other categorical and statistical data~\cite{dcn,deepfm}. Recognizing the effectiveness of user history~\cite{bst,nrms} in various recommendation scenarios, certain methods~\cite{din} explore the incorporation of a \textit{shallow} user encoder into CTR models, a strategy widely adopted in practice. It offers two key advantages over user ID representation in the context of large-scale industrial recommendation settings: 1) improved accuracy facilitated by the sequential features encoded by the user encoder, and 2) reduced memory usage by eliminating the need for a large user ID embedding table.

Meanwhile, the value of item contents, such as texts and images, has been recognized for providing more detailed and nuanced item representations compared to basic item IDs~\cite{yuan2023go,liu2022prec}. However, incorporating such item content information into CTR prediction models remains a challenge, especially within the
time and space constraints of industrial scenarios. 
The prevalent use of pretrained models 
across various domains
prompts the integrating of such models as end-to-end item encoders into CTR models, 
which we term as the \textbf{content-encoding paradigm}. Unfortunately, the user encoder requires the behavior sequence as input, 
necessitating the encoding of each item before fusing the behavior sequence to derive a user representation. This sequential dependency results in unacceptable training and inference inefficiency, impeding its adoption in industrial settings.


To address the efficiency challenge, the \textbf{embedding-based paradigm} opts to trade space for time. Some methods~\cite{wu2021newsbert,liu2022prec} utilize pretrained content encoders to convert item semantics into embedding vectors and cache them for subsequent CTR models. Furthermore, some methods~\cite{wu2022userbert,liu2022prec} explore the pretraining of user encoders, transforming user sequences into cached user embeddings. This approach effectively decouples item and user encoders via offline computing, leading to a substantial acceleration in both training and inference time. However, the use of pretrained models introduces a significant memory bottleneck; simply loading these extracted embeddings for training requires considerable memory.


\begin{figure*}[ht]
    \centering
    \includegraphics[width=.72\linewidth]{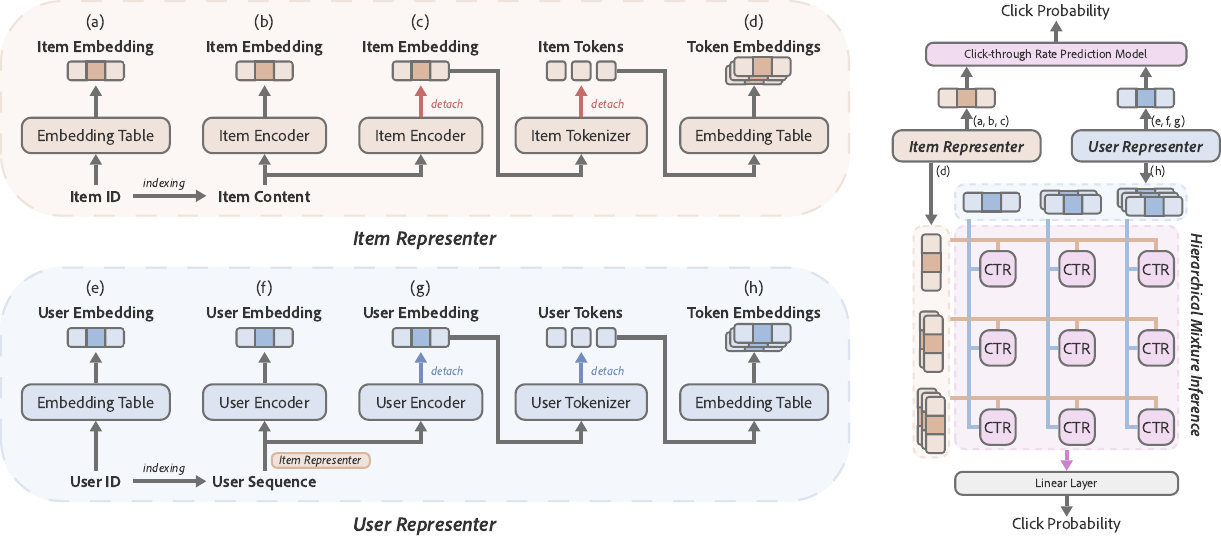}
    \caption{Our UIST framework for CTR prediction: featuring semantics-based item (d), and user (h) tokenizers, and a hierarchical mixture inference module.}
    \label{fig:overview} 
\end{figure*}

\definecolor{lightred}{RGB}{255, 200, 200}
\definecolor{lightyellow}{RGB}{255, 255, 200}
\definecolor{lightgreen}{RGB}{200, 255, 200}

\begin{table*}[]
\caption{Encoding paradigms for CTR prediction comparison. User and item representations correspond to the approaches in Fig.~\ref{fig:overview}. $V = 30,522$, $N = 100M$, and $M = 100M$ represent the vocabulary size of natural language, items, and users, respectively. $K = 4$ denotes the number of tokens used to represent an item or user; $d = 256$ and $D = 768$ denote the embedding dimensionality of the model and pretrained encoders. We use IST to represent single-layered semantic tokenization.
}
\label{tab:paradigm}

\resizebox{.72\linewidth}{!}{
\begin{tabular}{c|cc|c|c|cc|c}
\toprule
\textbf{Paradigm} & \textbf{Item Repr.} & \textbf{User Repr.} & \textbf{Efficiency} & \textbf{Examples} & \textbf{Item Memory} & \textbf{User Memory} & \textbf{Instantiation} \\
\midrule
Content-encoding        &(b)            & (f)            &  \cellcolor{lightred} $\times$          & -                 & V*d                & 0                  & 29.81M                  \\
\midrule
\multirow{2}{*}{\centering ID-based}          & (a)                  & (f)            &  \cellcolor{lightyellow} $\checkmark$       & DIN~\cite{din}               & N*d                & 0                  & 95.36G                 \\
\cmidrule{2-8}
& (a)                  & (e)                & \cellcolor{lightgreen} $\checkmark\checkmark$       & DCN~\cite{dcn}               & N*d                & M*d                & 190.72G                 \\
\midrule
\multirow{2}{*}{\centering Embedding-based}   & (c)             & (f)               & \cellcolor{lightyellow} $\checkmark$       & NP~\cite{liu2022prec}                & N*D                & 0                  & 286.10G                \\
\cmidrule{2-8}
& (c)             & (g)              & \cellcolor{lightgreen} $\checkmark\checkmark$       & PREC~\cite{liu2022prec}              & N*D                & M*D                & 572.20G                \\
\midrule
\multirow{2}{*}{\centering Semantic-token}    & (d)          & (f)               &  \cellcolor{lightyellow} $\checkmark$       & \smodel{} (Ours)               & N*K+K*d            & 0                  & 1.49G                  \\
\cmidrule{2-8}
& (d)          & (h)          & \cellcolor{lightgreen} $\checkmark\checkmark$       & \model{} (Ours)              & N*K+K*d            & M*K+K*d            & 2.98G                  \\
\bottomrule
\end{tabular}
}
\end{table*}

Recent work involves extracting item content features and condensing them into semantic IDs~\cite{tiger,jin2023language}, which efficiently capture the semantic representation of an item while maintaining its hierarchical structure. Building on this concept, we introduce a \textbf{semantic-token paradigm} and present \textbf{\model{}}, a \textbf{U}ser--\textbf{I}tem \textbf{S}emantic \textbf{T}okeniz\-ation approach that converts user sequences and item content into discrete user and item tokens, respectively. In contrast to the embedding-based paradigm, our approach greatly reduces space consumption while maintaining time efficiency, offering a substantial advantage for large-scale industrial recommender systems.

In Table~\ref{tab:paradigm}, we present a detailed efficiency and memory consumption comparison of such paradigms used for CTR prediction. Specifically, in the ``Instantiation'' column, we report the memory usage with vocabulary, user and item numbers set to reasonable settings for industrial applications. Compared to the embedding-based paradigm, our \model{} achieves a remarkable $\sim$200-fold (572.20G/2.98G) space compression, utilizing only four tokens.

Moreover, we devise a hierarchical mixture inference module to enhance the integration of hierarchical item and user tokens. This module dynamically adjusts the significance of various levels of granularity for user--item interactions. 


    

\section{User--Item Semantic Tokenization}


To balance time (model training and inference) and space (memory usage) efficiency, we introduce a user--item semantic tokenization framework, \model{}, following the semantic-token paradigm. \model{} comprises of three modules, illustrated in Figure~\ref{fig:overview}: two semantic tokenizers and one hierarchical mixture inference module. The semantic tokenizers initially transform dense and high-dimensional item and user embeddings into discrete tokens. Subsequently, the hyper controller, a hierarchical mixture inference (HMI) module, evaluates each user--item token pair using the base CTR modules and autonomously learns weights for each pair with supervision of the click labels. 

\subsection{Discrete Semantic Tokenization}
\label{sec:sentok}
Initially proposed for generative retrieval~\cite{tiger,jin2023language}, semantic tokenization has primarily been applied to summarize features on the item side. In this context, we introduce a unified pipeline that extends tokenization to both item and user sides, and we uniformly refer to both item content and user behavior as ``sequence''. It is important to note that tokenization of items precedes that of users, and during user tokenization, each item in user sequence is initialized with its corresponding representation extracted from the item tokenizer.

\textbf{Stage 1: Semantic Representation.} Traditional approaches often leverage pretrained language models like SentenceBERT to extract text-based content features on the item side by concatenating item attributes into a single sequence. However, this method is not directly applicable to the user side. Instead, we employ an autoencoder network for sequence representation learning.


Specifically, given a sequence $\mathbf{s}$ with length $L$, we initially transform it into an embedding sequence $\mathbf{E}^0$ through an embedding layer. Subsequently, a transformer~\cite{vaswani2017attention} encoder with $H$ layers learns the contextual knowledge of the sequence, expressed as:
$\mathbf{E}^h = \mathrm{ENC}^h ( \mathbf{E}^{h-1})$, where $h = 1, 2, \ldots, H.$
Next, an additive attention module~\cite{additive} merges the sequences into a unified representation, defined by:
$\mathbf{z} = \mathrm{ATTN}(\mathbf{E}^H).$
Following this, a transformer decoder reconstructs the original sequence. Comprising multiple decoder layers with causal attention, it integrates information from the sequence representation for autoregressive generation:
$\mathbf{D}^h = \mathrm{DEC}^h ( \mathbf{D}^{h-1}, \mathbf{z})$, where $h = 1, 2, \ldots, H,$
and the reconstruction loss is given by:
\begin{equation}
L_\text{sr} = -\frac{1}{L} \sum^L_{i=1} \log \left(\mathrm{G}\left(\mathbf{D}^H_i\right)_{\mathbf{s}[i]}\right),
\end{equation}
in which $\mathrm{G}$ represents the multi-layer perceptron classifier, and $\mathbf{s}[i]$ is the ground truth label for the output embedding $\mathbf{D}^H_i$. \\

\textbf{Stage 2: Discrete Tokenization.} 
The above enables random access to sequence embedding $\mathbf{z}$ for arbitrary user histories and item contents. 
Subsequently, we employ the residual quantization technique, RQ-VAE~\cite{rq}, to discretize the dense sequence representation into concise tokens. 
RQ-VAE is designed under an encoder-quantizer-decoder framework. From a macro perspective, the encoder maps the sequence embedding $\mathbf{z}$ to a latent vector by $\mathbf{v}$. The residual quantizer learns codes for the latent vector and sums up the code vectors ($\bar{\mathbf{v}}$) to approximate the latent vector $\mathbf{v}$. Finally, the decoder reconstruct the embedding $\bar{\mathbf{z}}$ from $\bar{\mathbf{v}}$.

Specifically, the residual quantizer operates iteratively as follows:
\textbf{i.} For each layer \(k\) (where \(k\) ranges from 1 to \(K\)), it maps the current vector \(\mathbf{v}^k\) to an index \(i^k\) by finding the nearest vector in the \(k\)-th layer's codebook, minimizing a distance function \(g\) (like Euclidean or Manhattan distance) between \(\mathbf{v}^k\) and each code vector \(\mathbf{C}^k_j\) in the codebook.
\textbf{ii.} The next vector \(\mathbf{v}^{k+1}\) is computed by subtracting the chosen code vector \(\mathbf{C}^k_{i^k}\) from \(\mathbf{v}^k\), thereby focusing on the residual (the part of the vector not yet quantized) for the next layer.
\textbf{iii.} The process begins with \(\mathbf{v}^1 = \mathbf{v}\), and iteratively refines the representation through the layers.

Formally, the above operations can be denoted as:
%
\begin{align}
f^k&: \mathbf{v}^k \rightarrow i^k, \quad \text{for } k \in {1, 2, \ldots, K},\\
\text{where }i^k &= \underset{j=1,\ldots,C}{\text{argmin }} g\left(\mathbf{v}^k, \mathbf{C}^k_j\right), \\
\mathbf{v}^{k+1} &= \mathbf{v}^k - \mathbf{C}^k_{i^k},~~
\text{and } \mathbf{v}^1 = \mathbf{v},
\end{align}
where $C$ is the codebook size for each layer.

Therefore, $(i^1, \ldots, i^K)$ is the tokenization result and the original vector $\mathbf{v}$ can be approximated by:
\begin{equation}
\mathbf{v} \approx \bar{\mathbf{v}} = \sum_k^K \mathbf{C}^k_{i^k}.
\end{equation}

The overall loss function $L_\text{et}$ is calculated as:
\begin{equation}
L_\text{et} = ||\mathbf{z} - \bar{\mathbf{z}}||^2 + \sum_k^K \left( | sg [ \mathbf{v}^k ] - \mathbf{C}^k_{i^k} |_2^2 + \beta | \mathbf{v}^k - sg [ \mathbf{C}^k_{i^k} ] |_2^2 \right),
\end{equation}
where $sg$ represents the stop gradient mechanism, and where the first term is the embedding reconstruction loss and the second term is the quantization loss. $\beta$ is the commitment cost that controls the influence of the vector movement.


\begin{table*}[ht]
\setlength\tabcolsep{3pt}

\caption{Performance comparison among different paradigms. We use red and green background to represent inefficient and efficient memory usage or inference latency, respectively.}
\label{tab:big-table}

\resizebox{.8\linewidth}{!}{
\begin{tabular}{r|c|c|c|ccc|ccc|ccc}
\toprule
& & \multicolumn{2}{c|}{\textbf{Memory}} & \multicolumn{3}{c|}{\textbf{DCN}} & \multicolumn{3}{c|}{\textbf{DeepFM}} & \multicolumn{3}{c}{\textbf{FinalMLP}} \\
\midrule
Paradigm & Repr.   & Item & User & AUC  & NDCG@5 & Latency & AUC   & NDCG@5  & Latency  & AUC   & NDCG@5   & Latency   \\
\midrule
Content-based & (b)/(f) & 30K $\times$ 256 & 0
& - & - & \cellcolor{lightred}66ms
& - & - & \cellcolor{lightred}65ms
& - & - & \cellcolor{lightred}68ms \\
\midrule
\multirow{2}{*}{ID-based}        & (a)/(f) & 65K $\times$ 256 & 0
& 0.5762 & 0.2808 & 12ms  
& 0.5665 & 0.2720 & 12ms
& 0.5777 & 0.2834 & 12ms \\
 & (a)/(e) & 65K $\times$ 256 & 94K $\times$ 256
& 0.5886 & 0.2862 & \cellcolor{lightgreen}3ms
& 0.5749 & 0.2781 & \cellcolor{lightgreen}3ms
& 0.5840 & 0.2833 & \cellcolor{lightgreen}3ms \\
\midrule
\multirow{2}{*}{Embedding-based} & (c)/(f) & \cellcolor{lightred}65K $\times$ 768 & 0
& 0.6301 & 0.3220 & 13ms
& 0.6195 & 0.3148 & 13ms
& 0.6259 & 0.3193 & 13ms \\
 & (c)/(g) & \cellcolor{lightred}65K $\times$ 768 & \cellcolor{lightred}94K $\times$ 768
& 0.6324 & 0.3247 & \cellcolor{lightgreen}3ms
& 0.6219 & 0.3161 & \cellcolor{lightgreen}3ms
& 0.6237 & 0.3174 & \cellcolor{lightgreen}3ms \\
\midrule
\multirow{2}{*}{Semantic-based}  & (d)/(f) \smodel{} & \cellcolor{lightgreen} 65K $\times$ 4 & 0
& 0.6297 & 0.3224 & 33ms
& 0.6183 & 0.3132 & 34ms
& 0.6241 & 0.3180 & 34ms \\
 & (d)/(h) \model{} & \cellcolor{lightgreen} 65K $\times$ 4 & \cellcolor{lightgreen} 94K $\times$ 4
& 0.6122 & 0.3085 & \cellcolor{lightgreen}3ms
& 0.6077 & 0.3028 & \cellcolor{lightgreen}3ms
& 0.6093 & 0.3050 & \cellcolor{lightgreen}3ms \\
\bottomrule
\end{tabular}
}
\end{table*}

\subsection{Hierarchical Mixture Inference (HMI)}

Following the semantic tokenization process, we obtain item and user tokens for each item $\mathbf{t}$ and user $\mathbf{u}$, represented as $(\mathbf{t}^1, \ldots, \mathbf{t}^K)$ and $(\mathbf{u}^1, \ldots, \mathbf{u}^K)$, respectively. For simplicity, we use the same number of layers $K$ during tokenization. Due to the nature of the residual quantization, these tokens are organized hierarchically, with the lower indices carrying primary component information. To effectively utilize user--item pairs at different levels, our hierarchical mixture inference module analyzes the contribution of each pair in the click-through rate prediction task.

Specifically, we transform one-hot tokens into dense embeddings through the user and item token embedding layers, denoted as $\mathbf{E}_{\mathbf{u}} \in \mathbb{R}^{K \times d}$ and $\mathbf{E}_{\mathbf{t}} \in \mathbb{R}^{K \times d}$. We then construct coarse-to-fine item and user embeddings based on the hierarchical tokens, defined as:
\begin{equation}
\bar{\mathbf{e}}^i_\mathbf{x} = \sum_j^i \mathbf{e}^k_\mathbf{x}, \quad i = 1, 2, \ldots, K, \text{and } \mathbf{x} \in \{\mathbf{u}, \mathbf{t}\},
\end{equation}
where $\mathbf{e}^i_\mathbf{x}$ is the $i$-th vector in $\mathbf{E}^i_\mathbf{x}$. For each user--item pair $(\bar{\mathbf{e}}^i_\mathbf{u}, \bar{\mathbf{e}}^j_\mathbf{t})$, we use a deep CTR model $\mathrm{M}$, such as DCN~\cite{dcn}, to predict click scores. Finally, a linear layer is employed to automatically weigh these scores. This yields the final click probability, formulated as:
\begin{equation}
    p = \mathrm{Linear}\left(\mathrm{M}\left(\bar{\mathbf{e}}^i_\mathbf{u}, \bar{\mathbf{e}}^j_\mathbf{t}\right)\right).
\end{equation}

As is standard, we use binary cross-entropy loss to train the recommendation task, calculated by:
\begin{equation}
    L_\text{rec} = - l \times \log \left(p\right) + (1 - l) \log \left(1 - p\right),
\end{equation}
where $l \in \{0, 1\}$ is the ground truth click label.\\

Furthermore, we also develop an \textbf{item-only semantic tokenization} (\textbf{\smodel{}}), tailored for the item side. By virtue of the shorter item tokens compared to the original item content, \smodel{} enhances both time and space efficiency in comparison to the content-encoding paradigm.
Nonetheless, when contrasted with \model{}, \smodel{} is a slower alternative due to its retention of the user encoder.
\begin{table}[]
\caption{Various aggregation mechanisms for user--item tokens. ``Add'' indicates the addition of item and user token embeddings to create a unified item and user representation. ``Layer'' signifies that only tokens from the same layer are input into the base CTR models. HMI represents our hierarchical mixture inference module.}
\label{tab:aggregation}
\resizebox{.65\linewidth}{!}{
\begin{tabular}{c|cc|cc}
\toprule
& \multicolumn{2}{c}{\textbf{DCN}} & \multicolumn{2}{c}{\textbf{DeepFM}} \\
\midrule
& AUC             & NDCG@5         & AUC              & NDCG@5           \\
\midrule
\textbf{Add}   & 0.5795          & 0.2836         & 0.5723           & 0.2774           \\
\textbf{Layer} & 0.6044          & 0.2963         & 0.5980           & 0.2929           \\
\textbf{HMI}   & 0.6122          & 0.3085         & 0.6077           & 0.3028          \\
\bottomrule
\end{tabular}
}
\end{table}

\section{Experiments}

We conduct offline experiments on a real-world news recommendation dataset, 
MIND~\cite{mind}, containing over 65K items and 94K users. We evaluate the effectiveness of our proposed \model{} against three modern deep CTR models:  DCN~\cite{dcn}, DeepFM~\cite{deepfm}, and FinalMLP~\cite{finalmlp}. We follow common practice~\cite{liu2022prec} to evaluate the recommendation effectiveness with AUC~\cite{auc} and nDCG~\cite{ndcg}. We also measure the inference time (latency) of each baseline for a single sample. During training, we employ the Adam optimizer~\cite{kingma2014adam} with a learning rate of 1e-3 for all paradigms. We set the number of transformer layers to 6 for 1) the item encoder (Fig.~\ref{fig:overview}b), 2) user encoder (Fig.~\ref{fig:overview}f), and 3) encoder--decoder used in semantic tokenization. For fair comparison, 
$\mathbf{z}$ in Section~\ref{sec:sentok} serves as the pretrained embeddings for the embedding-based paradigms. During semantic tokenization, we set the residual depth to 4 and the codebook size to 64. We will release the code and data for reproducible research\footnote{\url{https://github.com/Jyonn/SemanticTokenizer}}. 


Table~\ref{tab:big-table} compares the various paradigms across three deep CTR models, averaged over five independent runs. We make three key observations.  \textbf{i.} The content-based paradigm yields latencies exceeding 60ms, intolerable for industrial scenarios. \textbf{ii.} The single-layered ID-based and embedding-based approaches -- i.e., (a)/(f) and (c)/(f) pairings -- exhibit similar latency, as both include a user encoder to model user behavior; however, the performance of the embedding-based approaches are superior due to the use of the content-based item representation. On the other hand, the single-layered semantic-based \smodel{} approach is slightly slower because the item tokenization leads to a longer user sequence. \textbf{iii.} Our proposed \smodel{} and \model{} achieve substantial memory compression (approximately 200 times) compared to other paradigms, while maintaining up to 99\% accuracy (for \smodel{}) and 98\% accuracy (for \model{}) when compared to the state-of-the-art embedding-based paradigm. These findings, observed from diverse base models, validate the effectiveness and efficiency of our semantic-based approach. 

Table~\ref{tab:aggregation} examines various aggregation mechanisms for dual tokens, revealing that the simple addition and layer-wise approaches are inferior to our proposed HMI module.
\section{Conclusion}



We introduce a user--item semantic tokenization method, providing a streamlined approach to integrating item content into deep CTR models. Through our experimentation, we demonstrate the significant potential of semantic tokenization, initially proposed for generative retrieval, in boosting recommendation efficiency, particularly in industrial scenarios. 
Upon reflection, this method also offers new perspectives for applications such as dataset compression. We encourage researchers to further explore its potential.

\bibliographystyle{ACM-Reference-Format}
\bibliography{UIST}

\end{document}